\documentclass[fleqn,twoside,twocolumn,nofootinbib]{revtex4} 
\usepackage{ujp} 
\begin{document}
\title[Low-temperature dependences of the polarization]
{LOW-TEMPERATURE DEPENDENCES\\ OF THE POLARIZATION OF TERAHERTZ
EMISSION\\ FROM \boldmath$n$-GERMANIUM IN HEATING ELECTRIC FIELDS}%
\author{V.M. Bondar}
\affiliation{Institute of Physics, Nat. Acad. of Sci. of Ukraine}
\address{46, Nauky Ave., Kyiv 03680, Ukraine}
\email{ptomchuk@iop.kiev.ua}
\author{P.M. Tomchuk}%
\affiliation{Institute of Physics, Nat. Acad. of Sci. of Ukraine}%
\address{46, Nauky Ave., Kyiv 03680, Ukraine}%
\email{ptomchuk@iop.kiev.ua} \udk{} 
\razd{\secix}

\setcounter{page}{722}%
\maketitle

\begin{abstract}
As was noted in [3], in order to verify the assumption that the
behavior of the polarization dependences of the terahertz emission
by hot electrons from $n$-Ge is determined by the type of carrier
scattering, it is necessary to perform temperature measurements of
this dependence in the range between the temperatures, where
the scattering is determined by impurities and by acoustic lattice
vibrations, respectively. The given work presents the results of
such investigations.
\end{abstract}

\section{Introduction}

As was noted in [1--3], the terahertz emission by hot carriers  from
$n$-Ge has a specific feature -- it is polarized. Its polarization
dependences as functions of the angle between the polarization
vector and the heating field direction are periodic. The positions
of maxima and minima of these dependences can change due to
variations of the impurity concentration and the lattice
temperature. In addition, the polarization depends on the intensity
of a heating electric field and its direction relative to the
crystallographic axes, as well as on the degree of intervalley
repopulation. The reasons for a behavior of the polarization of this
emission were studied in a number of works [3--5]. It is known that
the main mechanisms of carrier scattering in $n$-Ge that determine
its electric characteristics are the impurity scattering at low
temperatures and the scattering by acoustic lattice vibrations at
high ones.\looseness=1

It was noted in [3] that, in order to verify the assumption that
different polarizations of the terahertz emission by hot carriers
in $n$-Ge in pure and doped materials are caused by different prevailing
scattering mechanisms, it is necessary to perform temperature
measurements of this phenomenon.

The idea of the method was to obtain the mentioned polarization
characteristics  at temperatures, at which each of these mechanisms
is determinative, and to follow the variation of these
characteristics under transition from one temperature region to the
other. In the case of low temperatures (impurity scattering), the
mobility of carriers in $n$-Ge changes with the temperature as $\mu
\sim T^{3/2}$, whereas it varies as $\mu \sim T^{-3/2}$ at high
temperatures (scattering by lattice vibrations). The maximum point
of the mobility, where the impurity scattering is replaced by the
scattering by acoustic phonons lies in the region $\sim $20 K for
pure $n$-Ge. One could expect that the polarization characteristic
will change its pattern passing through zero somewhere in this
temperature region. This will mean that the behavior of the
polarization dependence of the emission is determined by the type of
scattering of hot carriers. The experiments proposed and performed
in this work were to demonstrate whether it is true or
not.\looseness=1

\section{Experiments}

All measurements were carried out with the use of standard samples
produced by the typical technology [3]. Rectangular pulses of a
heating electric field had a length of 0.8 $mu$s, whereas their
amplitude could change in wide ranges. The detecting part of the
experimental set-up differed from the previous versions in the
arrangement of the filter, polarizer, detector, and emitting sample.
The main experimental difficulty of these measurements consists in
the fact that the $n$-Ge emitting sample, whose temperature should
be increased by an additional heating to 50 $\div $70 K, is located
near the Ge(Ga) semiconductor detector requiring helium temperature.
Such a temperature gradient at a distance of 8$\div $12 mm induces
very strong temperature instabilities and noise signals in the
detecting part, which makes measurements practically inadequate [3].
The problem was solved in the following way: the Ge(Ga) detector
immersed in helium and the emitting sample, whose temperature must
be changed from the helium one to 70$\div $80 K, were located at the
different ends of a vertical light guide. Locating the sample at
different distances from the detector (and the liquid-helium
surface), we can vary the temperature of the emitting sample within
the required limits. The necessity of an additional heating of the
sample disappeared, and temperature gradients were reduced to a
minimum. A diagram of the experimental set-up is presented in
Fig.~1.\looseness=1


\begin{figure}
\includegraphics[width=3.5cm]{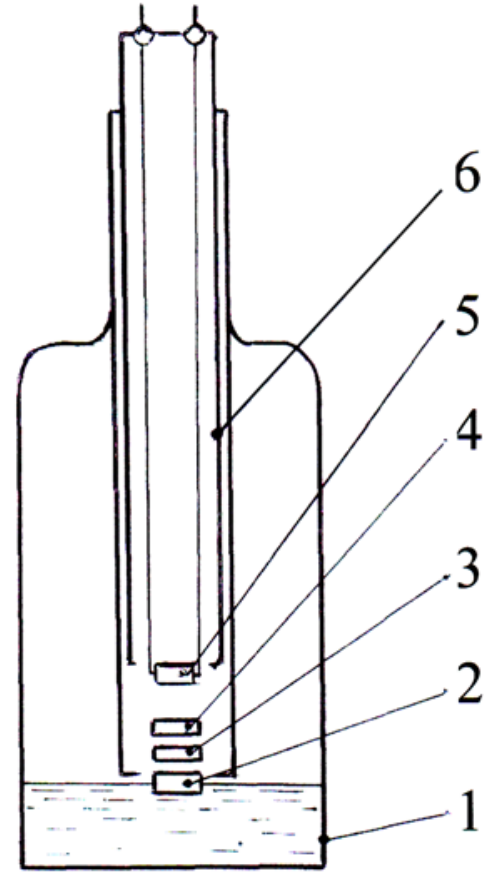}
\caption{ Diagram of the set-up used for temperature measurements of
the polarization of the terahertz emission from $n$-Ge: {\it 1} –-
helium cryostat; {\it 2} -- Ge(Ga) detector; {\it 3} -- filter; {\it
4} -– polarizer rotating in the horizontal plane; {\it 5} --
emitting sample; {\it 6} -- mobile light guide  }
\end{figure}

\section{Experimental Results and Their Discussion}

Figure 2 shows the results of measuring the temperature  dependence
of the terahertz emission of hot germanium carriers for a typical
GES-2.5 sample with the crystallographic orientation along the large
sample size $\langle 111\rangle$. One can see that, at the lowest
temperatures (6.6$\div $6.9) K, the behavior of the polarization
dependence is typical of doped samples. With increase in the sample
temperature, this dependence becomes straight, passes through zero
at 7.7 K, and then takes the form typical of a pure material. This
form does not change with the further increase of the temperature up
to 77 K. Thus, we observed an inflection point of the polarization
dependence close to 8 K. Relating this point to the transition from
the impurity scattering to the acoustic one (where the electron
mobility reaches a maximum), one can see that it differs from the
literature data by $\sim $10 K [6]. Additional experiments performed
for a number of samples with a different concentration yielded
similar results (Fig. 3). As it turned out later on, the difference
in the temperature measurements was due to the fact that the
semiconductor thermometer measured the temperature of the
environment surrounding the emitting sample, whereas the sample
itself was heated by pulses of the strong electric field ($\tau _i
=0.8$ $\mu$s, $\upsilon =$140 V, $\Im =2.6$ A,
$E_{\mathrm{rep}}$~=~6 Hz). Our calculations demonstrate that, at
these temperatures (taking into account the rapid decrease of the
heat capacity), the quantity of heat released in the sample (Fig. 2)
increases its temperature by 10 $\div $15 K, i.e. its real
temperature will be higher than that fixed by the thermometer by
$\sim  10 \div $15 K, which completely agrees with literature
data.\looseness=1


\begin{figure}
\includegraphics[width=8cm]{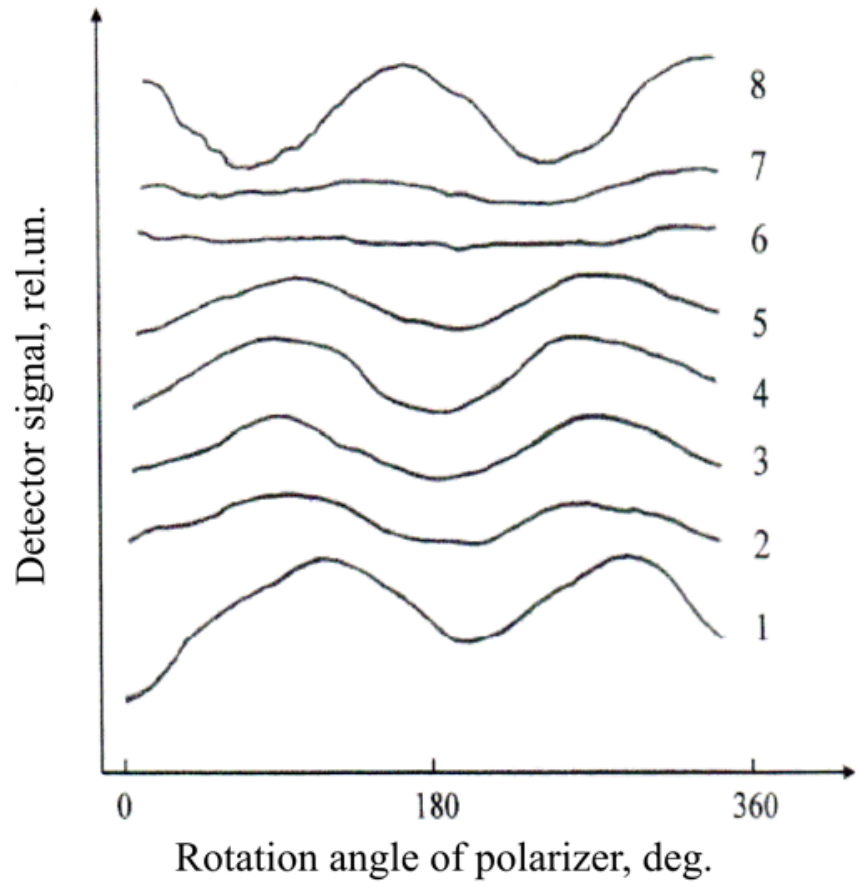}
\caption{Polarization dependence of the terahertz emission from
$n$-Ge (GES-2.5, $\langle 111 \rangle$) for different temperatures
(the curves are shifted in amplitude): {\it 1} --76 K; {\it 2} -- 57
K; {\it 3} -- 26 K; {\it 4} -- 14 K ; {\it 5} -- 8.8 K; {\it 6} --
7.7 K; {\it 7} -- 7.2 K; {\it 8} -- 6.6 K }
\end{figure}


\begin{figure}
\includegraphics[width=8cm]{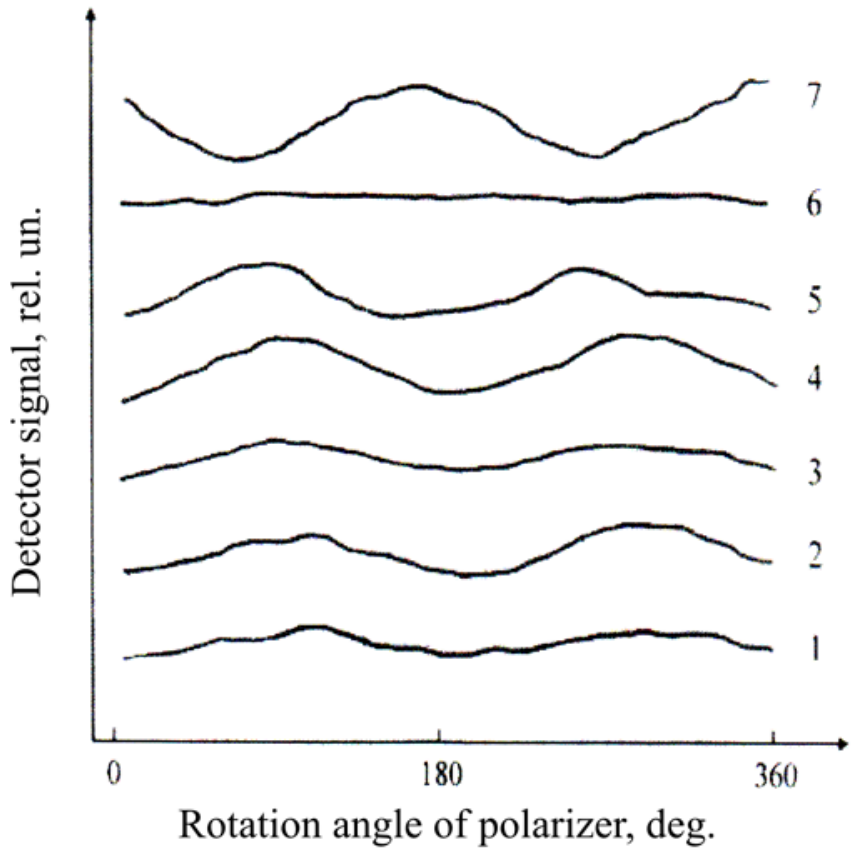}
\caption{Polarization dependence of the terahertz emission from
$n$-Ge (GES-0.3, $\langle 111 \rangle$) for different temperatures
(the curves are shifted in amplitude): {\it 1} -- 70 K; {\it 2} --
57 K; {\it 3} -- 26 K; {\it 4} -- 14 K; {\it 5} -- 7.2 K; {\it 6} --
6.4 K; {\it 7} -- 5.7 K }
\end{figure}

Hence, the refined position of the point of temperature  change of
the dominant scattering mechanism and the form of the curves given
in Figs. 2 and 3 allow us to make an unambiguous conclusion about
the relation between the angular polarization dependence and the
scattering mechanism. There took place a shift of the maxima
(minima): the regions of maxima in the polarization dependence in
the case of the impurity scattering changed to minima in the case of
the acoustic one.\looseness=1

Thus, the presented experimental results (Figs. 2 and 3) confirm the
proposed explanation of the behavior of the polarization
characteristic of the terahertz emission by hot electrons from
$n$-Ge depending on the type of carrier scattering.

\vskip3mm We thank O.G. Sarbey for the detailed discussion of the
results of this work and O.K. Florova for the given polarizers.

\rezume{%
ТЕМПЕРАТУРНІ~~~~~ ЗАЛЕЖНОСТІ~~~~~ ПОЛЯРИЗАЦІЇ \\ТЕРАГЕРЦОВОГО
ВИПРОМІНЮВАННЯ ГЕРМАНІЮ \\$n$-ТИПУ У ГРІЮЧИХ ЕЛЕКТРИЧНИХ ПОЛЯХ\\ В
ОБЛАСТІ НИЗЬКИХ ТЕМПЕРАТУР}{В.М. Бондар, П.М. Томчук} {Як було
відзначено в роботі [3] для правомірності припущення про те, що хід
поляризаційних залежностей терагерцового випромінювання гарячими
електронами з $n$-Ge визначається типом розсіювання носіїв,
необхідно провести температурні  виміри цієї залежності від
температур, де превалює розсіювання на домішках, до температур, де
превалює  розсіювання на акустичних коливаннях ґратки.  У даній
роботі наведено результати таких досліджень.}

\end{document}